\documentclass[preprintnumbers,twocolumn,superscriptaddress,aps,nofootinbib]{revtex4}
\usepackage{amssymb}
\usepackage[centertags]{amsmath}
\usepackage{txfonts}
\usepackage{epsfig}
\usepackage{bm}
\usepackage{color}
\usepackage{graphicx,graphics}
\usepackage{multirow}
\usepackage{float}
\usepackage[pdfstartview=FitH]{hyperref}
\hypersetup{colorlinks=true,citecolor=blue,linkcolor=blue,filecolor=black,urlcolor=blue}
\allowdisplaybreaks[2]

\begin{document}

\title{Baryon-anti-baryon Flavor Correlation in Quark Combination Models in Heavy Ion Collisions}

\author{Rui-qin Wang}
\affiliation{School of Physics, Shandong University, Jinan, Shandong 250100, China}
\affiliation{Key Laboratory of Particle Physics and Particle Irradiation (Shandong University), 
Ministry of Education, China}

\author{Feng-lan Shao}
\affiliation{Department of Physics, Qufu Normal University, Shandong 273165, China}

\author{Zuo-tang Liang}
\affiliation{School of Physics, Shandong University, Jinan, Shandong 250100, China}
\affiliation{Key Laboratory of Particle Physics and Particle Irradiation (Shandong University), 
Ministry of Education, China}

\begin{abstract}
 We extend the study of hadron yield correlation in combination models 
 in a recent publication to baryon-anti-baryon flavor correlations in heavy ion collisions. 
 We show that the universal behaviors of the anti-baryon to baryon yield ratios 
 as functions of that of $K^-$ to $K^+$ are naturally explained. 
 We study also the ``mixed ratios'' and propose other measurements that might be sensitive to the hadronization mechanism.
\end{abstract}

\maketitle

\section{introduction}  

Baryon-anti-baryon flavor correlation is usually regarded as a sensitive variable 
to test different hadronization models thus always draws much attention
in studying hadronization mechanism in different high energy reactions.  
Earlier studies can be found e.g. in $e^+e^-$ annihilations \cite{eeEXP1989ZPC,Liang1991PRD} 
and recently many discussions have been made in heavy ion collisions \cite{Zimanyi2000PLB,Shao2005PRC,thermal2009PLB}.
The most direct measurable quantity describing baryon-anti-baryon flavor correlation 
is the anti-baryon to baryon yield ratio $\langle N_{\bar B_i}\rangle/\langle N_{B_i}\rangle$ 
in a given kinematic region, where $B_i$ denotes different types of baryons,
and $\langle N_{h_i}\rangle$ denotes the average yield of the hadron $h_i$.
There are already quite abundant data available on yields of different $B_j$'s 
and the corresponding $\bar B_j$'s   
in heavy ion collision experiments at different energies for reactions using different nuclei, 
and from these data anti-baryon to baryon yield ratios 
have been \cite{PbarPK2003PRL,PbarPK2010PLB,PbarPK2011JPG,pikpPRL2012,PbarPK2002PRCNA44,
K20013062,plxoPRC62200,plxoPRCSPS,lxoPLB130,mixedRSTAR,Kp20arxiv,LXO-BES,BbB2013NPA,K2002PRCNA49,K30} 
and/or can be \cite{XiOm2013arxiv} calculated.                                           
It is interesting to see that the results seem to exhibit universal behaviors 
for $\langle N_{\bar B_i}\rangle/\langle N_{B_i}\rangle$ 
as functions of $\langle N_{K^-}\rangle/\langle N_{K^+}\rangle$ \cite{PbarPK2003PRL,PbarPK2010PLB,PbarPK2011JPG}.
 
In a recent paper \cite{HYC2012PRC}, we study the hadron yield correlations in heavy ion collisions. 
The hadrons are produced in the combination mechanism. 
The formalism is obtained by using the following two approximations and/or general assumptions:

 (1) Factorization: the flavor and momentum dependences of the combination kernel function are 
factorized and the momentum distributions of quarks and/or anti-quarks are taken as flavor independent;
 
 (2) Independent flavor production: 
the probability for production of different flavors in each of the new born 
quark-anti-quark pair in the reaction is taken as a constant independent from each other.  
Hence the numbers of different flavors of the newly produced quarks and anti-quarks 
follow the polynominal distribution with strangeness suppression factor $\lambda$.
The influence of net-quarks originated from the incident nuclei is taken into account by modifying 
the strangeness suppression factor for quark to 
$\lambda_q =2\lambda(1-\gamma_{net})/(2+\lambda\gamma_{net})$, 
 where $\gamma_{net}= \langle N_q^{net}\rangle/\langle N_q \rangle$, 
 $\langle N_q^{net}\rangle$ and $\langle N_q\rangle$ are 
the average numbers of net-quarks and quarks respectively .
 
Under these two assumptions, we have shown that there exist a set of relationships 
between the yields of different hadrons produced in the combination process. 
We presented the results for the case where only $J^P=0^-$ and $1^-$ mesons, 
$J^P=(1/2)^+$ and $(3/2)^+$ baryons are taken into account and 
showed that they are consistent with the data available. 
Apparently $\langle N_{\bar B_i}\rangle/\langle N_{B_i}\rangle$ is 
one special example of the hadron yield ratios describing hadron yield correlation 
studied in \cite{HYC2012PRC}. 
It is therefore natural to ask whether the data in particular the universal behaviors 
mentioned above can also be described by the formalism given there 
and whether there are more sensitive measurable quantities to describe 
the baryon-anti-baryon flavor correlation in heavy ion collisions. 
 
This note is intended to be an addendum to \cite{HYC2012PRC}.  
Here, we apply the formalism given in \cite{HYC2012PRC} to 
study the baryon-anti-baryon flavor correlation. 
We calculate $\langle N_{\bar B_i}\rangle/\langle N_{B_i}\rangle$ and propose other 
measurable quantities that can be used for the studies in this respect.

\section{$\bar B$ TO $B$ YIELD RATIOs} 

In \cite{HYC2012PRC}, we derived that, in the combination mechanism with the two approximations 
and/or assumptions mentioned above, 
hadron yields are given as functions of the strange suppression factor $\lambda$, 
the effective strangeness suppression factor $\lambda_q$, 
the average number of baryons $\langle N_{B}\rangle$ and that for mesons $\langle N_{M}\rangle$. 
In particular, the anti-baryon to baryon yield ratios are quite simple. 
They are the same for the directly produced ones and those in the final states 
including strong and electromagnetic decay contributions, 
and are given by \cite{HYC2012PRC},
\begin{align}
&\frac{\langle N_{\bar{p}}\rangle}{\langle N_{p}\rangle}
=\Bigl(\frac{2+\lambda_q}{2+\lambda}\Bigr)^{3}\frac{\langle N_{\bar{B}}\rangle}{\langle N_{B}\rangle},      
\label{pbarp} \\ 
&\frac{\langle N_{\bar{\Lambda}}\rangle}{\langle N_{\Lambda}\rangle}=\Bigl(\frac{2+\lambda_q}{2+\lambda}\Bigr)^{3}\frac{\lambda}{\lambda_q}\frac{\langle N_{\bar{B}}\rangle}{\langle N_{B}\rangle},\\ 
&\frac{\langle N_{\bar{\Xi}^{+}}\rangle}{\langle N_{\Xi^{-}}\rangle}
=\Bigl(\frac{2+\lambda_q}{2+\lambda}\Bigr)^{3}\Bigl(\frac{\lambda}{\lambda_q}\Bigr)^{2}\frac{\langle N_{\bar{B}}\rangle}{\langle N_{B}\rangle},    \\
&\frac{\langle N_{\bar{\Omega}^{+}}\rangle}{\langle N_{\Omega^{-}}\rangle}
=\Bigl(\frac{2+\lambda_q}{2+\lambda}\Bigr)^{3}\Bigl(\frac{\lambda}{\lambda_q}\Bigr)^{3}\frac{\langle N_{\bar{B}}\rangle}{\langle N_{B}\rangle},
\label{obaro}
\end{align}
Similarly, for kaons in the final state, we have, 
\begin{equation}
\frac{\langle N_{K^{-}}^f\rangle}{\langle N_{K^{+}}^f\rangle}=\frac{\lambda_q}{\lambda}
\frac{1+0.37\lambda}{1+0.37\lambda_q},        
\label{kmkp}
\end{equation} 
where the superscript $f$ denotes the results for hadrons in the final states. 
From these equations, we see that if there is no net-quark contribution, i.e., 
$\langle N_q^{net}\rangle$ or $\gamma_{net} =0$, 
we have $\lambda_q=\lambda$, $\langle N_{\bar{B}}\rangle=\langle N_{B}\rangle$, 
and hence all the ratios shown by Eqs.~(\ref{pbarp}-\ref{kmkp}) are unity. 
We also see that, at a given $\lambda$, $\langle N_{K^-}^f\rangle/\langle N_{K^{+}}^f\rangle$ 
is a singly valued decreasing function of $\gamma_{net}$. 
We insert $\lambda_q =2\lambda(1-\gamma_{net})/(2+\lambda\gamma_{net})$ 
into Eq. (\ref{kmkp}) and obtain, 
\begin{equation}
\frac{\langle N_{K^{-}}^f\rangle}{\langle N_{K^{+}}^f\rangle}=
(1-\gamma_{net}) \frac{1+0.37\lambda}{1+0.37\lambda+0.13\lambda\gamma_{net}}. 
\label{kmkpvsgamma}
\end{equation}
Here, we not only clearly see that ${\langle N_{K^{-}}^f\rangle}/{\langle N_{K^{+}}^f\rangle}$ is 
a singly valued function of $\gamma_{net}$ but also see that, for small $\gamma_{net}$, 
${\langle N_{K^{-}}^f\rangle}/{\langle N_{K^{+}}^f\rangle}$ is basically nothing else but 
$(1-\gamma_{net})$. 

\begin{figure}[htbp]
\centering
 \includegraphics[width=0.65\linewidth]{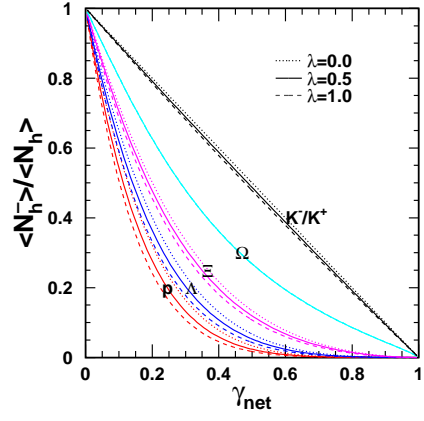}\\
 \caption{(Color online) Anti-hadron-to-hadron yield ratios as functions of 
$\gamma_{net}=\langle N_q^{net}\rangle/\langle N_q\rangle$ 
 for different values of $\lambda$.}
 \label{fig:BbarBvsGamma}
\end{figure}

Also, from Eqs. (\ref{pbarp}-\ref{obaro}), we see that, 
besides $\langle N_{\bar B}\rangle/\langle N_B\rangle$,   
the yield ratios of $\bar B_j$ to $B_j$ are also functions of $\gamma_{net}$.  
It can also be expected that $\langle N_{\bar B}\rangle/\langle N_B\rangle$ 
should also be a function of $\gamma_{net}$. 
In fact, in \cite{HYC2012PRC}, we derived that, for reasonably large $N_q$ ($>100$, say), 
in combination mechanism, the ratio of the average number of anti-baryons
to mesons produced in the process can be parameterized with good approximation by,
\begin{equation} 
\frac{\langle N_{\bar B}\rangle}{\langle N_M\rangle}=\frac{1}{12}(1-\gamma_{net})^{2.8}, 
\label{eq:BbarM}
\end{equation}
and this leads to,
\begin{equation}
\frac{\langle N_{\bar B}\rangle}{\langle N_B\rangle}
=\frac{(1-\gamma_{net})^{3.8}} {(1-\gamma_{net})^{2.8}+4\gamma_{net}}. 
\label{eq:BbarB}
\end{equation}
We insert Eq. (\ref{eq:BbarB}) into Eqs. (\ref{pbarp}-\ref{obaro}) and obtain,
\begin{align}
&\frac{\langle N_{\bar{p}}\rangle}{\langle N_{p}\rangle}
=\Bigl(\frac{2}{2+\lambda\gamma_{net}}\Bigr)^{3}\frac{(1-\gamma_{net})^{3.8}}{(1-\gamma_{net})^{2.8}+4\gamma_{net}},      
\label{pbarpvsgamma} \\ 
&\frac{\langle N_{\bar{\Lambda}}\rangle}{\langle N_{\Lambda}\rangle}=\Bigl(\frac{2}{2+\lambda\gamma_{net}}\Bigr)^{2}\frac{(1-\gamma_{net})^{2.8}}{(1-\gamma_{net})^{2.8}+4\gamma_{net}},\\ 
&\frac{\langle N_{\bar{\Xi}^{+}}\rangle}{\langle N_{\Xi^{-}}\rangle}
=\Bigl(\frac{2}{2+\lambda\gamma_{net}}\Bigr)\frac{(1-\gamma_{net})^{1.8}}{(1-\gamma_{net})^{2.8}+4\gamma_{net}},    \\
&\frac{\langle N_{\bar{\Omega}^{+}}\rangle}{\langle N_{\Omega^{-}}\rangle}
=\frac{(1-\gamma_{net})^{0.8}}{(1-\gamma_{net})^{2.8}+4\gamma_{net}}.
 \label{obarovsgamma}
\end{align}
We see that they are functions of $\gamma_{net}$. 
In Fig.~\ref{fig:BbarBvsGamma}, we present the results of these ratios as functions of $\gamma_{net}$. 
During the calculations, we also find out that the results are not very sensitive to 
the value of the strangeness suppression factor $\lambda$ but there are some influences
in particular for $\bar p$ to $p$ ratio. 
In Fig.~\ref{fig:BbarBvsGamma}, we show besides the results obtained 
at $\lambda=0.5$ that is usually used in heavy ion collisions 
and also two limiting cases for $\lambda=1$ (no strange suppression) and the maximum 
strange suppression by taking the limit at $\lambda\to 0$ for comparison.

We see that all these ratios are indeed monotonically decreasing functions of $\gamma_{net}$.  
We see in particular that $\langle N_{K^-}^f\rangle/\langle N_{K^{+}}^f\rangle$ deviates little from 
$1-\gamma_{net}$ which just corresponds to the limit at $\lambda=0$ shown in the figure. 
Because of this, we can simply use $\langle N_{K^-}^f\rangle/\langle N_{K^{+}}^f\rangle$ 
to replace $\gamma_{net}$ and express the anti-baryon to baryon ratios as functions of 
$\langle N_{K^-}^f\rangle/\langle N_{K^{+}}^f\rangle$. 
We emphasize that these functions should be universal in the sense that they are 
consequences of the combination mechanism with the assumptions of factorization and 
flavor independent production mentioned in the introduction of this note. 
They should be independent of the nuclei used in the collisions, 
independent of the reaction energies.
Furthermore, they should also be the same in different kinematic regions, and
also the same for different centralities as long as the combination mechanism is at work. 

\begin{figure}[htbp]
\centering
 \includegraphics[width=0.85\linewidth]{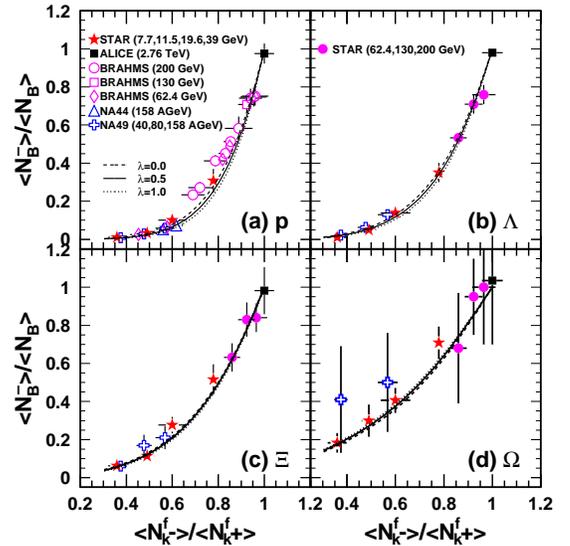}\\
 \caption{(Color online) Anti-baryon-to-baryon yield ratios versus 
 that of $K^-$ to $K^+$ compared with the data from different experiments \cite{PbarPK2003PRL,PbarPK2010PLB,PbarPK2011JPG,pikpPRL2012,PbarPK2002PRCNA44,
 K20013062,plxoPRC62200,plxoPRCSPS,lxoPLB130,Kp20arxiv,LXO-BES,K2002PRCNA49,XiOm2013arxiv,BbB2013NPA}.}
 \label{fig:BbarB-K}
\end{figure}

In Fig.~\ref{fig:BbarB-K}, we show the results compared with the 
data obtained by different  collaborations. 
These data are taken from Refs.~\cite{PbarPK2003PRL,PbarPK2010PLB,PbarPK2011JPG,pikpPRL2012,PbarPK2002PRCNA44,
K20013062,plxoPRC62200,plxoPRCSPS,lxoPLB130,Kp20arxiv,LXO-BES,K2002PRCNA49,XiOm2013arxiv,BbB2013NPA}. 
Those from STAR and NA49 are all for central rapidity regions such as $-0.5<y<0.5$,  
while those from NA44 are for rapidity regions around $y\sim2$ to $4$, 
and those from BRAHMS are for different rapidity intervals in the whole rapidity regions.
We see that the results are indeed in good agreement with the data 
available \cite{PbarPK2003PRL,PbarPK2010PLB,PbarPK2011JPG,pikpPRL2012,PbarPK2002PRCNA44,
K20013062,plxoPRC62200,plxoPRCSPS,lxoPLB130,Kp20arxiv,LXO-BES,K2002PRCNA49,XiOm2013arxiv,BbB2013NPA}. 

Furthermore, from Eqs.~(\ref{pbarp}-\ref{obaro}), 
we also see clearly that, if we build the ratio between any two 
${\langle N_{\bar B_i}\rangle/\langle N_{B_i}\rangle}$'s from them, 
we obtain a very simple result, i.e., a power of $\lambda/\lambda_q$. 
For example, we have,
\begin{align}
&\frac{\langle N_{\bar{\Lambda}}\rangle/\langle N_{\Lambda}\rangle}{\langle N_{\bar{p}}\rangle/\langle N_{p}\rangle}
=\frac{\langle N_{\bar{\Xi}^{+}}\rangle/\langle N_{\Xi^{-}}\rangle}{\langle N_{\bar{\Lambda}}\rangle/\langle N_{\Lambda}\rangle}
=\frac{\langle N_{\bar{\Omega}^{+}}\rangle/\langle N_{\Omega^{-}}\rangle}{\langle N_{\bar{\Xi}^{+}}\rangle/\langle N_{\Xi^{-}}\rangle}        \nonumber    \\
& =\Bigl(\frac{\langle N_{\bar{\Xi}^{+}}\rangle/\langle N_{\Xi^{-}}\rangle}{\langle N_{\bar{p}}\rangle/\langle N_{p}\rangle}\Bigr)^{0.5}=\frac{\lambda}{\lambda_q}\approx\frac{\langle N_{K^{+}}^f\rangle}{\langle N_{K^{-}}^f\rangle}.   \label{QCT-F} 
\end{align} 
They are valid for both directly produced baryons and those in the final states.  
We emphasize that these relationships are more intrinsic in the combination mechanism 
since they are independent of e.g. the parameterization given by Eq.~(\ref{eq:BbarM}). 
They could provide more sensitive tests to the hadronization mechanism. 

\begin{figure}[htbp]
\centering
 \includegraphics[width=0.8\linewidth]{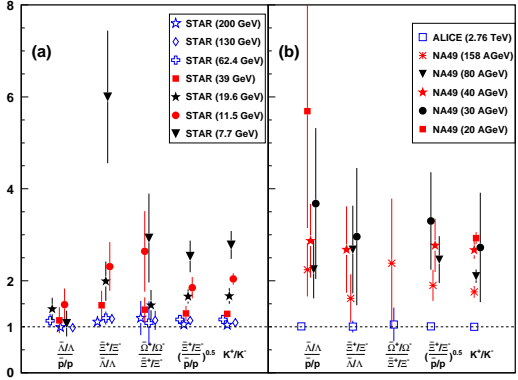}\\
 \caption{(Color online) Mixed hadron yield ratios derived from the data of yields of 
 the corresponding hadrons obtained in different experiments \cite{K20013062,plxoPRC62200,PbarPK2011JPG,Kp20arxiv,LXO-BES,lxoPLB130,mixedRSTAR,
 BbB2013NPA,XiOm2013arxiv,pikpPRL2012,K2002PRCNA49,K30,plxoPRCSPS}. 
 See the text for more details. 
 The horizontal position is only used to distinguish different ratios.}
 \label{MixR}
\end{figure}

Such kinds of ``mixed ratios" of yields of hadrons have been analysed 
in experiments such as given in \cite{lxoPLB130,mixedRSTAR} where results on
$(\langle N_{\bar{\Lambda}}\rangle/\langle N_{\Lambda}\rangle)/(\langle N_{\bar{p}}\rangle/\langle N_{p}\rangle)$,
$(\langle N_{\bar{\Xi}^{+}}\rangle/\langle N_{\Xi^{-}}\rangle)/(\langle N_{\bar{\Lambda}}\rangle/\langle N_{\Lambda}\rangle)$,
$(\langle N_{\bar{\Omega}^{+}}\rangle/\langle N_{\Omega^{-}}\rangle)/(\langle N_{\bar{\Xi}^{+}}\rangle/\langle N_{\Xi^{-}}\rangle)$,
and $[(\langle N_{\bar{\Xi}^{+}}\rangle/\langle N_{\Xi^{-}}\rangle)/
(\langle N_{\bar{p}}\rangle/\langle N_{p}\rangle)]^{0.5}$
at $\sqrt{s_{NN}}=130$ GeV derived from the measurements by STAR Collaboration are given. 
We plot them in Fig.~\ref{MixR} (a). 
Similar hadron yield ratios have also been given by STAR Collaboration with energy scan \cite{K20013062,plxoPRC62200,PbarPK2011JPG,Kp20arxiv,LXO-BES}. 
We calculate the corresponding ratios using them and show the results in the same figure. 
Also, there are data at LHC and SPS energies \cite{BbB2013NPA,XiOm2013arxiv,pikpPRL2012,K2002PRCNA49,K30,plxoPRCSPS}. 
We show the results derived from them in Fig.~\ref{MixR} (b). 
All the data are for different energies but all in the central rapidity regions.

From Fig.~\ref{MixR} (a) and (b), we see that, at relatively high energies such as 
those at high RHIC and LHC, these ratios indeed fall into one straight horizontal line respectively, 
consistent with the results given by Eq.~(\ref{QCT-F}). 
However, at lower energies, in particular, at $\sqrt{s_{NN}}=7.7$ GeV 
and $E_{beam}=20$ AGeV, there seems indications that the 
results deviate from those expected in the quark combination mechanism 
with the two assumptions and/or approximations mentioned in the introduction part of this note.
This might be taken as a signature of the failure of the quark combination mechanism in this energy range.
More precise measurements are needed to clarify this.

\section{$B\bar B$ FLAVOR CORRELATION FACTORS} 

Another quantity to measure the baryon-anti-baryon flavor correlation can be the 
correlation factor $C_{B_j \bar B_k}$ defined as, 
\begin{equation}
 C_{B_j \bar B_k} \equiv \frac{\langle N_{B_j \bar B_k} \rangle}{\langle N_{B_j} \rangle \langle N_{\bar B_k} \rangle},  
 \label{eqdefi:CBjBbark}
\end{equation}
where $N_{B_j \bar B_k}=N_{B_j} N_{\bar B_k}$ is 
the number of all possible $B_j \bar B_k$ pairs in the system considered. 
Clearly, if the flavor of anti-baryon and that of baryon are completely uncorrelated, we have, 
\begin{equation}
C_{B_j \bar B_k}|_{uncorr}=1.
\end{equation}
For the case that the flavor of $\bar B_j$ is completely correlated with that of $B_j$, we have,
$N_{B_j}=N_{\bar B_j}$ and
 \begin{equation}
C_{B_j \bar B_j}|_{totcorr}=1+D^2_{B_j}/\langle N_{B_j} \rangle ^2,
\end{equation}
where $D^2_{B_j}\equiv\langle N_{B_j}^2 \rangle-\langle N_{B_j} \rangle ^2$ is the dispersion 
of the distribution of $N_{B_j}$. 
For example, if $N_{B_j}$ distributes as a Poissonian, we 
have $D^2_{B_j}=\langle N_{B_j}\rangle$, 
thus $C_{B_j \bar B_j}|_{totcorr}=1+1/\langle N_{B_j} \rangle$;
if $N_{B_j}$ distributes as a negative binomial, we have, 
$D^2_{B_j}=\langle N_{B_j}\rangle+\langle N_{B_j}\rangle^2/k_j$, and 
$C_{B_j \bar B_j}|_{totcorr}=1+1/k_j+1/\langle N_{B_j} \rangle$, 
where $k_j$ is a parameter in the negative binomial distribution. 
  
In the combination model as formulated in \cite{HYC2012PRC} with the two approximations 
and/or assumptions of factorization and independent flavor production mentioned 
in the introduction part of this note, $C_{B_j \bar B_k}$ can be easily calculated. 
The result is given by,  
 \begin{equation}
 C_{B_j \bar B_k} = \frac{\langle N_{B \bar B} \rangle}  {\langle N_{B} \rangle \langle N_{\bar B} \rangle}.  \label{eqdefi:CBjBbark}
\end{equation}
They are all the same for different combinations of $B_j$ and $\bar B_k$. 
If we scale $C_{B_j \bar B_k}$ by ${\langle N_{B \bar B} \rangle} /{\langle N_{B} \rangle \langle N_{\bar B} \rangle}$ and denote it as 
$c_{B_j \bar B_k}\equiv C_{B_j \bar B_k}/({\langle N_{B \bar B} \rangle} /{\langle N_{B} \rangle \langle N_{\bar B} \rangle})$, 
we have $c_{B_j \bar B_k}=1$ independent of $B_j$ and $\bar B_k$.
This is a very clear prediction of the combination models that can be tested 
by future experiments.

\begin{figure}[htbp]
\centering
 \includegraphics[width=0.65\linewidth]{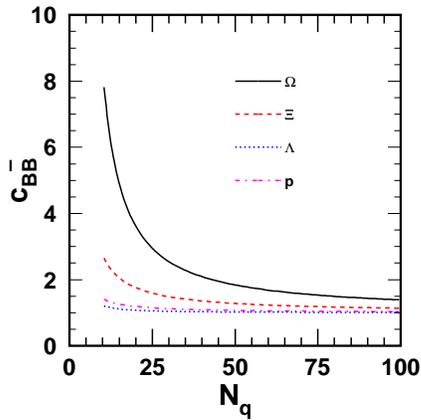}\\
 \caption{(Color online) Baryon-anti-baryon flavor correlation factors $c_{B_j \bar B_k}$, 
 scaled by ${\langle N_{B\bar B} \rangle }/{\langle N_{B} \rangle \langle N_{\bar B} \rangle}$, 
 as a function of $N_q$ in the case that the flavor is globally compensated.}
 \label{fig:CBB}
\end{figure}

One effect that might influence the flavor correlation factor $c_{B_j\bar B_k}$ 
in a system is the global flavor compensation. 
To make an estimation, we consider a system of equal number $N_q$ of quarks and 
anti-quarks and suppose the flavor is compensated globally. 
The expressions showing the influence are not difficult to derive but the 
results are a little bit lengthy. 
The simplest one is for $ {\Omega^- \bar \Omega^+}$, where we have, 
\begin{align}
c_{\Omega^- \bar \Omega^+}&
 = 
 [ (N_q-3)( N_q-4)(N_q -5)p_s^3 +9(N_q-3)(N_q -4)p_s^2 \nonumber\\
 &~~~ + 18(N_q-3) p_s+6]/N_q(N_q-1)(N_q -2)p_s^3, 
\end{align} 
where $p_s=\lambda/(2+\lambda)$.
To get a feeling of the size of such effect, we show in Fig.~\ref{fig:CBB} the numeric results 
of $c_{B_j\bar B_{k=j}}$ for different types of baryons 
scaled by ${\langle N_{B\bar B} \rangle }/{\langle N_{B} \rangle \langle N_{\bar B} \rangle}$ 
as a function of $N_q$ when global flavor compensation is taken into account. 
We see that there are indeed significant influences at not very large $N_q$ in particular for $\Omega$. 
The influences are small for large $N_q$. 
Usually, when we consider a sub-system of hadrons in a given kinematic region, this influence 
is expected to be small since the global flavor compensation only applies to the whole system 
produced in the collision. 

\section{summary}
In summary, we study the baryon-anti-baryon flavor correlation in quark combination in heavy ion collisions. 
We adopt the formalism derived in \cite{HYC2012PRC} by making the assumptions of 
flavor-momentum-factorization and independent flavor production at the quark level. 
We show that the observed universal behaviors of anti-baryon to baryon yield ratios 
as functions of that of $K^-$ to $K^+$ are naturally explained. 
We propose further measurements in this connection and make predictions for future experiments.  

The authors would like to thank Q. B. Xie for helpful discussions.
This work is supported in part by the Major State Basic Research Development Program 
in China (No. 2014CB845400) 
and the National Natural Science Foundation of China 
projects (Nos. 11035003, 11375104 and 11175104).

\end{document}